\documentclass[reprint,amsmath, superscriptaddress, amssymb, aps, longbibliography, prl]{revtex4-1}
\usepackage{graphicx}
\usepackage{dcolumn}
\usepackage{bm}
\usepackage[T1]{fontenc}
\usepackage[utf8]{inputenc}
\usepackage{physics}
\usepackage{xcolor}
\newcommand\hl[1]{%
  \bgroup
  \hskip0pt\color{blue}%
  #1%
  \egroup
}
\DeclareMathAlphabet{\pazocal}{OMS}{zplm}{m}{n}

\begin{document}
\linespread{1}

\title{Exciton transport in a moiré potential: from hopping to dispersive regime}
\author{Willy Knorr}
\email{knorrw@uni-marburg.de}
\author{Samuel Brem}
\author{Giuseppe Meneghini}
\author{Ermin Malic}
\affiliation{Department of Physics, Philipps University, 35037 Marburg, Germany}
\begin{abstract}
   The propagation of excitons in TMD monolayers has been intensively studied revealing interesting many-particle effects, such as halo formation and non-classical diffusion. Initial studies have investigated how exciton transport changes in twisted TMD bilayers, including Coulomb repulsion and Hubbard-like exciton hopping. In this work, we investigate the twist-angle-dependent transition of the hopping regime to the dispersive regime of effectively free excitons. Based on a microscopic approach for excitons in the presence of a moiré potential, we show that the hopping regime occurs up to an angle of approximately $2^{\circ}$ and is well described by the Hubbard model. At large angles, however, the Hubbard model fails due to increasingly delocalized exciton states. Here, the quantum mechanical dispersion of free particles with an effective mass determines the propagation of excitons.  Overall, our work provides microscopic insights into the character of exciton propagation in twisted van der Waals heterostructures.     \\    

\end{abstract}
\maketitle
\section{Introduction}

Recently, vertically stacked monolayers of transition metal dichalcogenides (TMDs) into van der Waals heterostructures have been used to study many-body phenomena including charge carrier dynamics and charge transport properties\cite{wang2018colloquium, mueller2018exciton,schaibley2016valleytronics,brem2022terahertz}. In a TMD monolayer, exciton properties can be broadly tuned by strain and dielectric engineering \cite{niehues2018strain,aslan2018strain,latini2015excitons,rosati2021dark}. Here, in particular, propagation of excitons has been intensively studied, including interesting effects, such as the formation of spatial rings (halos) \cite{kulig2018exciton,perea2019exciton} as well as non-classical diffusion \cite{wagner2021nonclassical} or negative diffusion \cite{rosati2020negative,berghuis2021effective}. In this context, van der Waals heterostructures open up a variety of interesting research questions due to their rich exciton energy landscape that is tunable by adding a twist angle  \cite{yu2017moire, merkl2019ultrafast,brem2020tunable,shabani2021deep}. In a twisted heterostructure, one finds a superposition of the respective geometries of the individual monolayers generating a moiré superlattice \cite{andrei2021marvels,tran2019evidence,jin2019observation,alexeev2019resonantly,forg2021moire}. Similar to a classical moiré pattern, this superlattice is strongly twist-angle dependent, leading to different energy landscapes including  trapped exciton states \cite{baek2020highly,seyler2019signatures}. Consequently, lattices of moir\'e-trapped interlayer excitons in TMD heterobilayers have been suggested to realize bosonic Hubbard systems \cite{wu2018hubbard,tang2020simulation,gotting2022moir} and transport measurements have demonstrated that the diffusion of excitons is strongly altered by the moir\'e potential \cite{yuan2020twist,choi2020moire}.  \\
One of the most promising application potentials of twisted heterostructures lies within the tunability of the excitonic transport properties, ranging from trapped exciton states appearing at very small twist angles and quasi free delocalized exciton states at larger angles exhibiting a  drastically different transport behavior \cite{brem2020tunable}.  In this work, we investigate the different exciton transport regimes in the presence of a moiré potential in a twisted MoSe$_2$/WSe$_2$ heterostructure. 
\begin{figure}[t!]
\centering
\begin{minipage}{.37\textwidth}
  \centering
  \includegraphics[width=\textwidth]{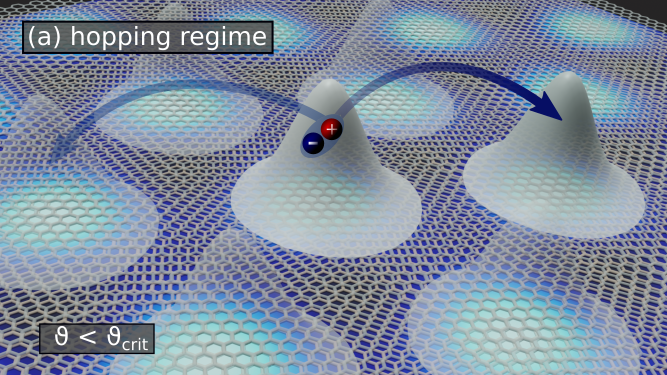}
  \label{fig:sub1}
\end{minipage}%
\hfill
\begin{minipage}{.37\textwidth}
  \centering
  \includegraphics[width=\textwidth]{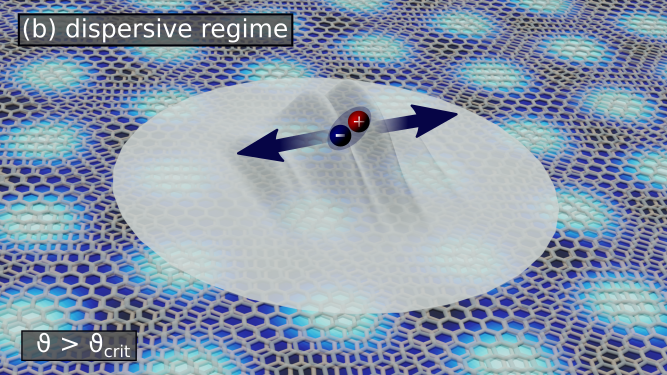}
  \label{fig:sub2}
\end{minipage}
\caption{Schematic illustration of exciton transport in a moiré potential created by a twisted heterostructure. For different twist angles $\vartheta$ we encounter different transport regimes. (a) At small angles, we observe a Hubbard hopping-driven transport (b) while at large twist angles we reach a dispersive regime.}
\label{fig:schematic}
\end{figure}
Based on a microscopic approach towards the excitonic eigenstates within a moiré potential\cite{brem2020tunable}, we compute the free propagation of a spatially inhomogeneous exciton distribution by applying different approximation frameworks. In particular, we consider the equations of motion for a moiré exciton wave packet by taking into account (i) the next-neighbor hopping between moiré super-cells and (ii) the propagation expected within an effective mass approximation. To decide about the suitability of the chosen approximation we compare these results to a numerical solution obtained by taking into account the full topology of moiré mini-bands. We find that the next-neighbor hopping corresponding to the Hubbard model is well suited for the regime of small twist angles up to approximately $2^{\circ}$, cf. Fig. \ref{fig:schematic}(a). In contrast, at large twist angles, excitons propagate like free particles and the moiré potential acts as a small perturbation. Here, the Hubbard model fails and excitons are found to propagate quasi-free through the heterostructure with a modified mass, cf. Fig.~\ref{fig:schematic}(b). Finally, we have investigated the change in the exciton mass due to the superlattice potential. We find that for $\vartheta>4^{\circ}$ the effective mass converges towards the  free exciton mass. For decreasing angles, however, the exciton mass is enhanced by almost one order of magnitude at $\vartheta=1^{\circ}$.

\section{theory}
In this work we focus on the free propagation of excitons within a moiré potential, i.e. balistic transport, without considering the effects of phonon- or defect-assisted scattering and neglecting density-dependent effects, such as exciton-exciton interaction. 
First, we set up an excitonic Hamilton operator in the low density regime and derive an exact numerical solution for the scattering-free propagation of moir\'e excitons. For large twist angles, the moiré mini-band structure converges into the free exciton dispersion, such that moiré excitons are expected to propagate like quasi-free particles in a lattice potential. To verify this, we  derive the time evolution of quasi-free excitons. On the other hand, the flat band structure at small twist angles indicates a strongly reduced exciton propagation, suggesting hopping-dominated transport. Here, we introduce an excitonic Hubbard model and calculate the exciton occupation numbers as function of time and space. Finally, we compare all these solutions as a function of the twist angle.\\
\\
\textbf{Moiré Excitons.} In atomically thin TMD heterostructures electrons and holes are tightly bound as excitons due to the strong attractive Coulomb interaction \cite{wang2018colloquium}. In the presence of a twist angle,  a  moiré superlattice occurs, which strongly influences the exciton center-of-mass (CoM) motion. Since excitations in TMDs are strongly dominated by excitons, we express the Hamilton operator in an exciton basis\cite{haug1984electron,katsch2018theory}. The latter is represented by the excitonic creation and annihilation operators  $X^{\dagger}_{\mu\mathbf{Q}}/X_{\mu\mathbf{Q}} $ \cite{brem2020tunable} and we find
\begin{align}\label{Eq:singleHamiltonian_exciton}
    H=\sum_{\mu\mathbf{Q}}\mathcal{E}_{\mathbf{Q}}^{\mu}X^{\dagger}_{\mu\mathbf{Q}}X_{\mu\mathbf{Q}}+\sum_{\mu\mathbf{Q}\mathbf{q}}\mathcal{M}_{\mathbf{q}}^{\mu}X^{\dagger}_{\mu\mathbf{Q}+\mathbf{q}}X_{\mu\mathbf{Q}}.
\end{align}
The Hamiltonian includes the free part with the exciton dispersion $\mathcal{E}_{\mathbf{Q}}^{\mu}$ and the moiré potential with the matrix element $\mathcal{M}_{\mathbf{q}}^{\mu}$. More details can be found in the supporting information. Here,  $\mathbf{Q}$ represents the CoM momentum, $\mu$ refers to the exciton states, which can be either composed of electron-hole pairs in the same layer (intralayer exciton) or in different layers (interlayer excitons) \cite{merkl2019ultrafast}. Furthermore, $\mathbf{q}$ represents the momentum transfer during the interaction with the moiré potential. Note that we restrict our considerations to the K point, where the electronic wave functions are strongly localized in one of the two layers. Since we focus on the MoSe$_2$/WSe$_2$ heterostructure, where the KK interlayer exciton is the lowest energy states \cite{lu2019modulated, brem2020tunable}, we can also neglect the hybridization between intra- and interlayer exciton states \cite{gillen2018interlayer,brem2020hybridized}.  Exploiting the periodicity of the moiré potential, we apply a zone-folding approach \cite{brem2020tunable,brem2020hybridized,brem2022terahertz}, where every point outside of the mini-Brillouin zone is mapped back inside with a unique reciprocal lattice vector $\mathbf{G}$ of the superlattice. Thus, the band structure is given by a series of moiré exciton subbands $\nu$. In the moiré exciton basis, the Hamiltonian becomes diagonal yielding
\begin{align}\label{Eq_single_particle_final}
    H = \sum_{\nu\mathbf{Q}}E_{\mathbf{Q}}^{\nu}Y^{\dagger\nu}_{\mathbf{Q}}Y^{\nu}_{\mathbf{Q}},\quad Y^{\dagger\nu}_{\mathbf{Q}} = \sum_sc^{*\nu}_{s}(\mathbf{Q})X^{\dagger}_{\mathbf{Q}+\mathbf{G}_s},
\end{align}
where $Y^{\nu}_{\mathbf{Q}}$ represents the moiré exciton operator. Here, $E_{\mathbf{Q}}^{\nu}$ and $c^{*\nu}_{s}(\mathbf{Q})$ are the eigenvalues and the eigenstates in the new basis, respectively. More details can be found in the supplementary information. \\

\textbf{Exact Solution.}
Having determined the excitonic eigenstates within the moiré potential, the time dependent real-space wave function for a given initial distribution can be computed as 
\begin{align}\label{Eq:exact_solution}
    \psi(\mathbf{r},t)=\sum_{\mathbf{Q},\nu}\tilde{\psi}_{\nu}(\mathbf{Q},0)\chi_{\mathbf{Q},\nu}(\mathbf{r})\,\text{exp}\Bigg(-\frac{i}{\hbar}E_{\mathbf{Q}}^{\nu}t\Bigg)
\end{align}
with the Bloch wavefunction $\chi_{\mathbf{Q},\nu}=\sum_sc_s^{\nu}(\mathbf{Q})\text{exp}(-i(\mathbf{Q}+\mathbf{G}_s)\mathbf{r}$ and the projection of the initial state $\tilde{\psi}_{\nu}(\mathbf{Q},0)=\braket{\nu,\mathbf{Q}}{\psi(0)}$. In experiments, the excitation is typically realized by Gaussian-shaped laser pulses determining the shape of the initial exciton distribution. Here, the real space width is on the order of the excitation wavelength and is therefore in the $\mu$m-range \cite{kulig2018exciton}. Thus, the initial width of the wave function is large compared to the moiré lattice vector, that is $<100$nm for the twist angles $>1^\circ$ considered in this work \cite{huang2022excitons}. As a result, the width of the initial wave function in momentum space becomes very narrow compared to the moiré mini-Brillouin zone. Consequently, we perform a Taylor expansion of the twist angle dependent band structure around the $\gamma$-point and obtain isotropic and parabolic bands with $E_{Q}\approx\hbar^2 Q^2/(2m_{\text{eff}})$ with the effective mass $1/m_{\text{eff}}=1/\hbar^2\partial^2E_{Q}/\partial Q^2$. Within this quadratic approximation we can solve the integral in Eq.\eqref{Eq:exact_solution} analytically. We obtain the exciton density distribution as
\begin{equation}\label{Eq:density_free}
    \rho(\mathbf{r},t) = \frac{1}{2\pi\sigma^2(t)}\text{exp}\left(-\frac{\mathbf{r}^2}{2\sigma^2(t)}\right),
\end{equation}
with the time-dependent variance $\sigma^2(t)=\sigma_0^2\left(1+4\hbar^2t^2/(m_{\text{eff}}^2\sigma_0^4)\right)$, where $\sigma_0^2$ represents the variance of the initial density distribution. Instead of Eq. \ref{Eq:density_free} we use the full band structure $E_\mathbf{Q}$ and the exact expression for the wave function in Eq. \ref{Eq:exact_solution} to numerically compute the time evolution of the wave packet. By comparing the results with Eq. \ref{Eq:density_free} we can then extract an effective mass for the propagation of moiré excitons in experimentally relevant situations.
\\

 \textbf{Bose Hubbard Model.} At very small angles, however, the exciton band structure is flat \cite{brem2020tunable} suggesting a hopping-dominated exciton transport. For this reason, we introduce the Bose-Hubbard model, which is a powerful tool for calculating the dynamics of strongly localized hopping-driven states. In the past it was already successfully implemented to describe the dynamics of bosonic atoms in optical lattices \cite{jaksch1998cold,jaksch2005cold} as well as quantum phase transitions from superfluid to Mott insulators \cite{jaksch1998cold,greiner2002quantum}. We introduce an excitonic Bose-Hubbard model by exploiting the already introduced exciton Hamiltonian in Eq.\eqref{Eq_single_particle_final}, while transforming the operators into a Wannier basis $b^{\dagger}_{\nu,n} = 1/\sqrt{N}\sum_{\mathbf{Q}}\text{exp}(i\mathbf{Q}\cdot\mathbf{R}_n)Y^{\nu\dagger}_{\mathbf{Q}}$, with $\mathbf{R}_n$ as the n-th moiré superlattice minimum. In this basis the Hamilton operator reads
\begin{equation}\label{Eq:Hopping_Hamiltonian}
    H = \sum_{n,m,\nu} t_{n,m}^{\nu} b^{\dagger}_{\nu,n}b_{\nu,m},
\end{equation}
with $t_{n,m}^{\nu}$ as the hopping term, which can be calculated via the overlap of the Wannier wavefunctions $W_{n}(\mathbf{r})=1/\sqrt{N}\sum_{\mathbf{Q},s}e^{-i\mathbf{Q}\cdot\mathbf{R}_n}c_s^{\nu}(\mathbf{Q})\text{exp}\{-i\mathbf{Q}+\mathbf{G}_s)\mathbf{r}\}$ at different lattice positions n and m, yielding
\begingroup
\begin{align*}\label{Eq:Hopping_term}
    t^{\nu}_{n,m} &= \int d^2r~W_n^*(\mathbf{r})HW_m(\mathbf{r})=\frac{1}{N}\sum_{\mathbf{Q}}e^{i\mathbf{Q}\cdot(\mathbf{R}_m-\mathbf{R}_n)}E_{\mathbf{Q}}^{\nu},
\end{align*}
\endgroup
with the twist-angle dependent band structure $E_{\mathbf{Q}}^{\nu}$. As we will see, the topology of the band structure gives direct information about the hopping term: (i) Flat bands representing localized states lead to small hopping terms due to the small overlap of the Wannier orbitals. (ii) Parabolic bands representing delocalized states lead to larger hopping terms reflecting a larger orbital overlap. The hopping term  is an exact representation of the moiré eigenstates in Wannier basis. In the Hubbard model the hopping is usually restricted to the lowest orbital and  hopping between the nearest neighbors ($t_{n,m}\approx 0$ for $|\mathbf{R}_m-\mathbf{R}_n|>a_m$).

Within the approximation of the Hubbard Model we calculate the spatiotemporal exciton dynamics. For this purpose, we define the exciton density for the lowest exciton subband $\nu=0$ in the Wannier basis
\begin{equation}
    \rho(\textbf{r},t)=\langle\psi^{\dagger}(\textbf{r})\psi(\textbf{r})\rangle = \sum_{n,m}W^*_m(\textbf{r})W_n(\textbf{r})\rho_{nm}(t),
\end{equation}
where the field operators are expressed as $\psi^{\dagger}(\textbf{r})=\sum_nW_n(\textbf{r})b_n^{\dagger}$. Here,  $\rho_{nm}=\langle b^{\dagger}_{n}b_{m}\rangle$ represents the density matrix, which drives the temporal evolution of the exciton density and which can be calculated via the Heisenberg equation of motion
\begin{align}
    \dot{\rho}_{nm}=-\frac{i}{\hbar}\sum_i\left(t_{mi}\rho_{ni}-t_{in}\rho_{im}\right),
\end{align}
which we solve numerically via a Runge-Kutta algorithm. Thus, we have developed two different approaches to calculate the exciton density distribution. To quantify the transport behavior for trapped and delocalized excitons as well as the intermediate transport regime, we use the definition of variance as the second central moment $\sigma^2(t)=\int d^2r\mathbf{r}^2\rho(\mathbf{r},t)$ of the distribution. We find that in all  approximations presented above, the dispersion length defined as $\xi(t)=\sqrt{\sigma^2(t)-\sigma_0^2}$ increases linearly with time. We thus define a constant of motion $\alpha$, which quantifies the propagation and which we refer to as dispersion parameter
\begin{align}\label{Eq:alpha}
    \alpha = \frac{\sigma_0^2}{2}\partial_t \xi(t).
\end{align}
For the case of a parabolic dispersion (cf. Eq. (\ref{Eq:density_free})), it holds $\alpha=\hbar/m_{\text{eff}}$ allowing to unambiguously determine the effective mass $m_{\text{eff}}$ as function of the twist angle. In order to obtain comparable results for the Hubbard model and the exact solution, we choose initial conditions, in which only the lowest moiré subband is occupied, i.e. a superposition of ground state Wannier orbitals with a Gaussian envelope. Hence, the Hubbard simulation is initialized via $\rho_{nm}(t=0)=\delta_{nm} \mathcal{G}(\sigma_0,|\mathbf{R}_n-\mathbf{R}_0|)$ and the exact simulation via the corresponding projection $\tilde{\psi}_\nu(\mathbf{Q},0)=\delta_{\nu,0}\int d^2r\chi^\ast_{\mathbf{Q},\nu}(\mathbf{r})\sqrt{\mathcal{G}(\sigma_0,r)}$, where $\mathcal{G}(\sigma_0,r)$ is the Gaussian envelope with the width $\sigma_0$. 

\section{Exciton Transport Regimes}
\begin{figure}[t!]
    \includegraphics[width=0.5\textwidth]{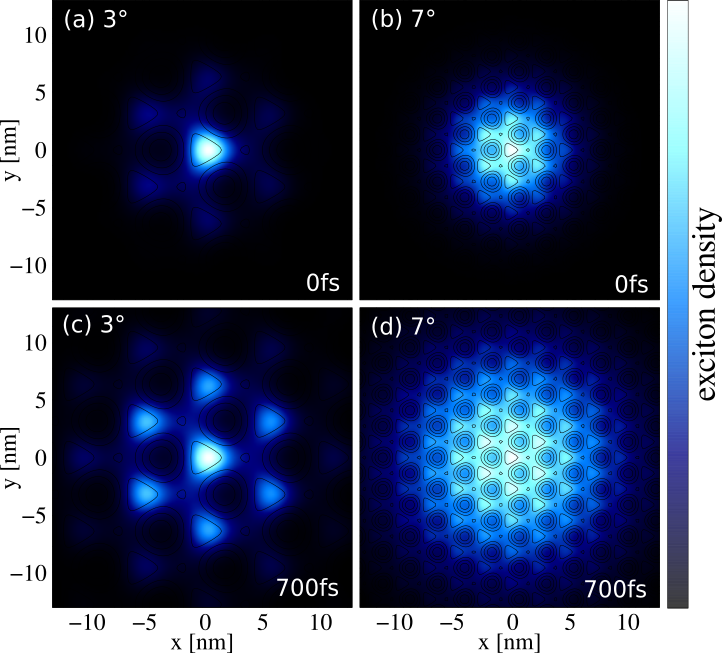}
    \caption{Exciton propagation in a  twisted $\text{MoSe}_2/\text{WSe}_2$ heterostructure. (a)-(b) The initial exciton density distribution at $3^{\circ}$ and $7^{\circ}$, respectively. The contour lines represent the moiré potential. (c)-(d) Exciton distribution after 700 fs. For smaller twist angles excitons are strongly localized leading to a suppressed propagation compared to larger angles, where the Wannier orbitals have a larger overlap. }
    \label{fig:dynamic}
\end{figure}

In the following, we analyze the distinct transport regimes in TMD heterostructures exhibiting a twist-angle tunable moiré potential. Figure \ref{fig:dynamic} shows the spatiotemporal exciton dynamics determined by the exact expression in Eq. (\ref{Eq:exact_solution}) for two different twist angles. The corresponding moiré potentials are represented by the contour lines in the background, illustrating the decrease in the supercell size when increasing the twist angle. 
In the case of a small twist angle of $\vartheta=3^{\circ}$ we observe that only a few moir\'e sites are initially occupied (at time 0 fs) and that these sites are almost isolated and well localized. In contrast, for a larger twist angle of $\vartheta=7^{\circ}$ with the same initial Gaussian, a much larger number of sites are excited representing a more delocalized exciton distribution. After $t=700$fs, we observe a clear broadening of the exciton distribution for both twist angles. However, we also find that the exciton transport behavior differs drastically. While the exciton occupation remains localized at single sites for the smaller twist angle, excitons propagate in all directions for the larger angle.  
To quantify the propagation rate, we have introduced the dispersion parameter $\alpha$ allowing us to obtain deeper insights into different characteristics of the distinct exciton transport regimes.\\

\textbf{Dispersive and hopping regime.} 
In the previous section we introduced the excitonic Bose-Hubbard model and the corresponding hopping term $t^{\nu}_{n,m}$. 
\begin{figure}[t!]
    \centering
    \includegraphics[width=0.48\textwidth]{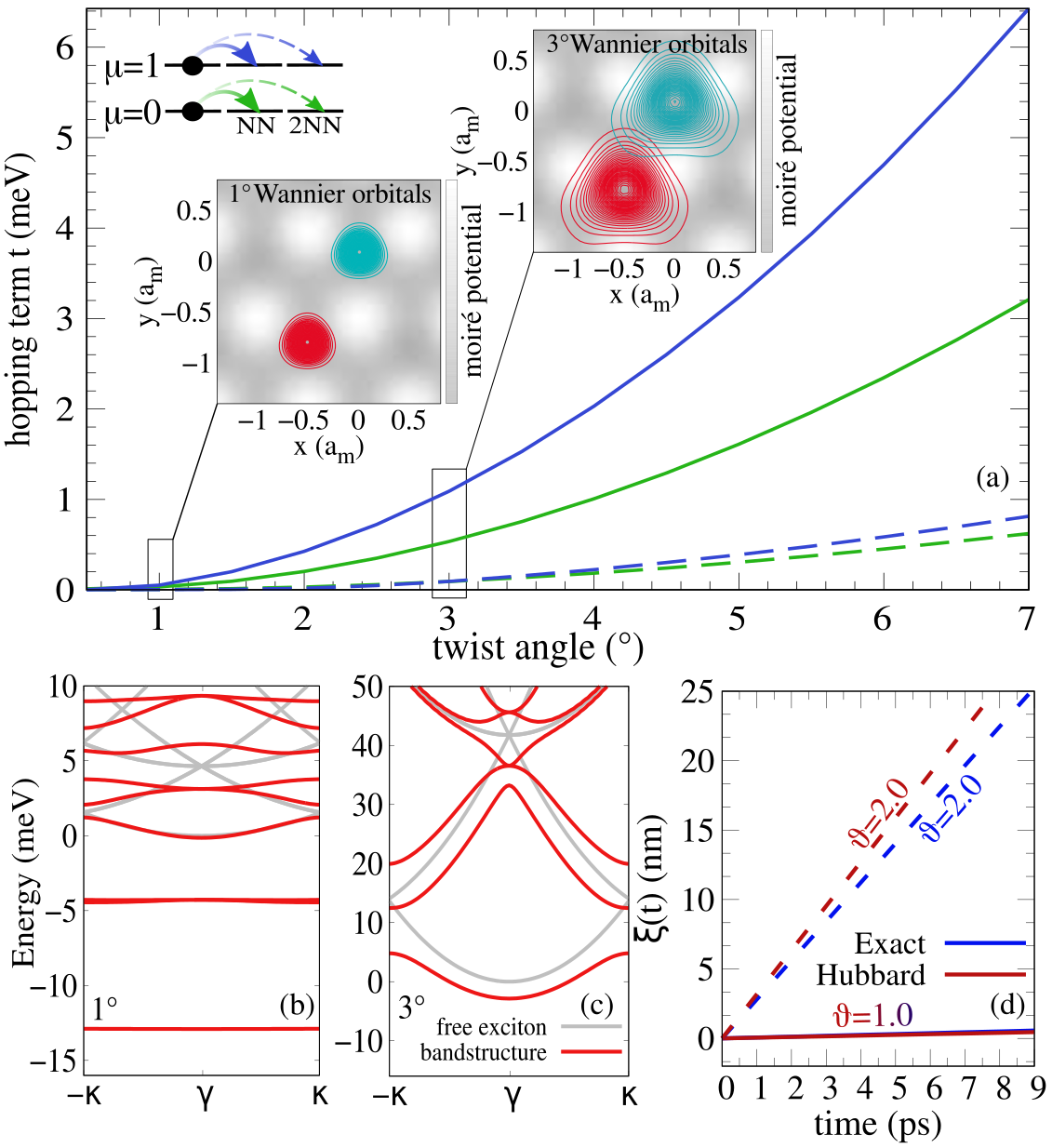}
    \caption{(a) Hopping term over the twist angle $\vartheta$ for nearest neighbor (NN) and next-neighbor (2NN) interaction. We  depict the hopping term for the ground and first excited exciton states, i.e. $\nu=0,1$. Due to the more pronounced overlap between the Wannier wavefunction (insets in (a)) the nearest neighbor interaction is the dominant term for the hopping interaction. (b)-(c) Moiré exciton subbands for interlayer excitons in the twisted $\text{MoSe}_2/\text{WSe}_2$ heterostructure at $\vartheta=3^{\circ}$ and $\vartheta=1^{\circ}$, respectively. (d) Time evolution of the dispersion length for the exact solution (blue) in comparison with the Hubbard model (red) for two different twist angles. For the lower angle of $\vartheta=1.0^{\circ}$, the length is the same, while already for $\vartheta=2^{\circ}$ we find clear differences.}
    \label{fig:Hopping}
\end{figure}
Figure \ref{fig:Hopping}(a) illustrates the hopping term as a function of the twist angle $\vartheta$ for the  ground state $\nu=0$  (green lines) as well as for the excited state $\nu=1$ (blue lines) and taking into account the nearest (solid lines) and next-nearest (dashed lines) neighbor interactions. 
For small $\vartheta$, the hopping term converges to zero as here excitons are completely trapped and there is no overlap of their orbital functions and thus no transport. The moiré superlattice constant increases for small twist angles resulting in a large potential barrier for tunneling between neighboring sites. 
In contrast, at large twist angles, the hopping term monotonously  increases reflecting the increasing delocalization of the Wannier orbitals and thus resulting in a larger orbital overlap, cf. the insets of Fig.~\ref{fig:Hopping}(a) for $\vartheta=1^{\circ}$ and $\vartheta=3^{\circ}$ illustrating the orbitals of two neighboring sites within the moiré potential. This general trend can also be understood from the twist-angle evolution of the moiré subbands. In Fig.~\ref{fig:Hopping}(b) and (c) we show the exemplary band structures for $\vartheta=1^{\circ}$ as well as $\vartheta=3^{\circ}$. We observe a flat band structure for $1^\circ$ corresponding to negligible group velocity, whereas for $3^\circ$ the bands exhibit a parabolic dispersion indicating a larger mobility. 

The most striking difference between between nearest- (NN) and next-neighbor (2NN) hopping (solid vs dashed lines) is the slope of their increase as a function of the twist angle, cf. Fig.~\ref{fig:Hopping}(a). The hopping due to the nearest-neighbor interaction increases much faster  due to the larger overlap of directly adjacent sites. 
Furthermore, the 2NN hopping becomes only important at larger twist-angles, while it is completely negligible for angles smaller than $\vartheta<2^{\circ}$.
Regarding the hopping term for different exciton subbands, we find the same general behavior for the lowest and first excited state (green and blue lines). However, the latter exhibits a more efficient  hopping as excited exciton states are less strongly bound and have a more delocalized wave functions and thus a larger overlap.

\begin{figure}[t!]
    \centering
    \includegraphics[width=0.48\textwidth]{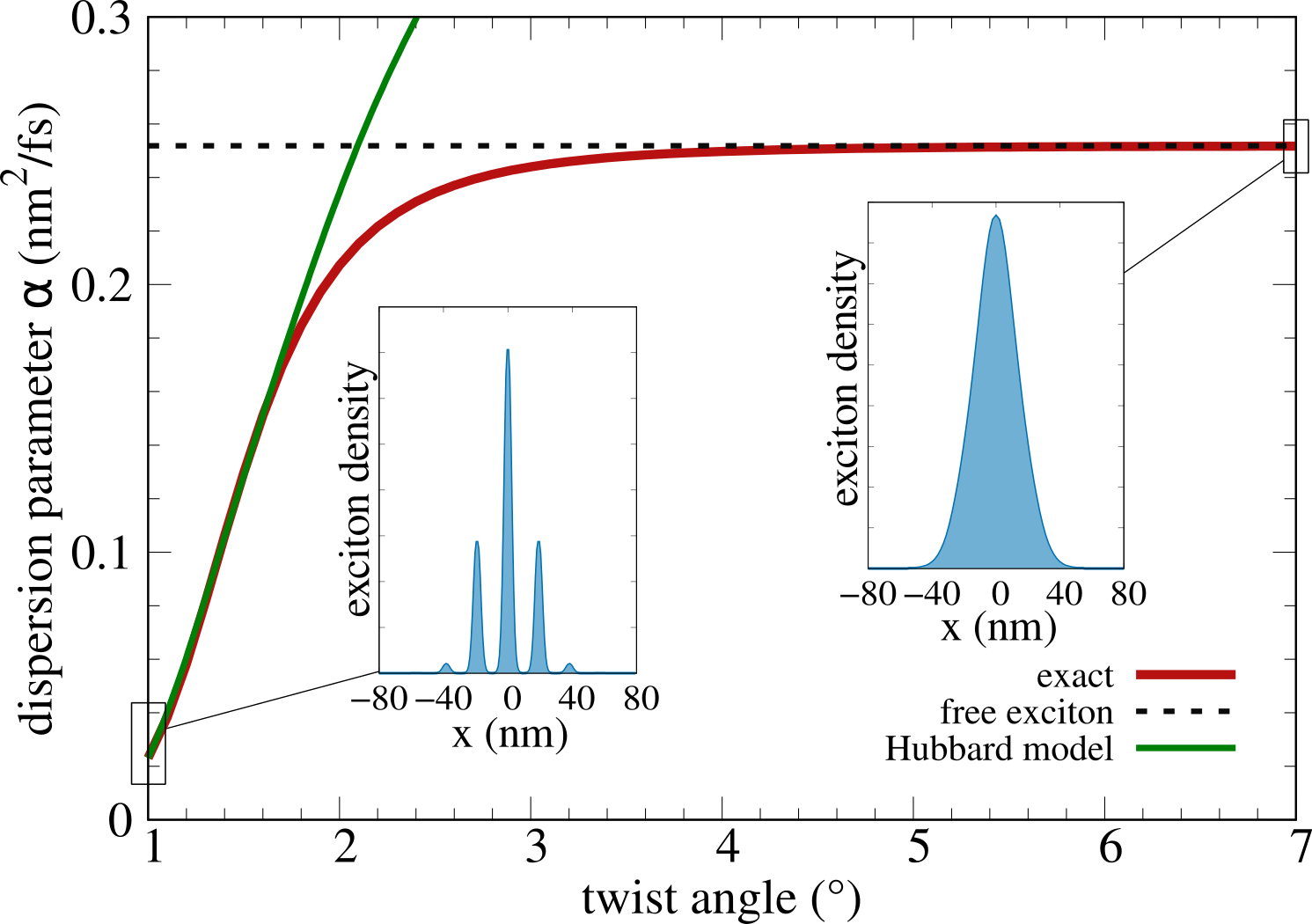}
    \caption{Dispersion parameter $\alpha$ (Eq.\eqref{Eq:alpha}) as a function of the twist angle $\vartheta$ showing a direct comparison of the  exact numerical solution for the propagation of excitons in a moiré potential (red line) with the Hubbard and the free-exciton solutions. The exact solution converges to the free solution (dashes line) for $\vartheta>4^\circ$ and describes the dispersive regime that is characterized by a delocalized exciton distribution (inset at 7$^\circ$). The exact solution goes over to the Hubbard solution for $\vartheta>1.8^\circ$ that is characterized by a localized exciton distribution (inset at $1^{\circ}$). }
    \label{fig:alpha}
\end{figure}

In Fig.~\ref{fig:Hopping}(d), we additionally show the dispersion length $\xi(t)$ as defined above as function of time for the Hubbard model as well as for the exact solution.  In both cases we observe a linear behavior in time, which means that in both cases the slope of the curve, i.e. the dispersion parameter $\alpha$ (from Eq. (\ref{Eq:alpha})) can be used to quantify the transport behavior. In this figure, we also find that for a small twist angle of $\vartheta=1^\circ$, the dispersion length for the Hubbard model and the exact solution overlap, indicating that the Hubbard model is accurate. However, already at $\vartheta=2^\circ$, a clear deviation is observed indicating that the Hubbard model becomes quickly insufficient to describe exciton propagation within the moiré potential at small twist angles.

The dispersion parameter $\alpha$ is a constant of motion and quantifies the propagation velocity. This allows us to investigate and compare the efficiency of exciton transport in different regimes. Figure \ref{fig:alpha} shows  $\alpha$ as a function of the twist angle $\vartheta$ as extracted from the exact solution (red solid), the solution for the completely free excitons (dashed) and  the Hubbard solution (green line). We observe that the dispersion parameter $\alpha$ rapidly increases up to a value of about 0.25 nm$^2$/fs that is reached at a twist angle of about 4$^\circ$. For further increasing angles, $\alpha$ remains constant reflecting the constant dispersion parameter for free excitons (dashed line). Here, the exciton distribution is delocalized over a large number of supercells well resembling the effective mass limit, cf. the inset for 7$^\circ$ in Fig. \ref{fig:alpha}.
Exciton propagation in a moiré potential with a large twist angle resembles the dispersion of a quantum-mechanic wave packet. Thus we denote this transport regime as dispersive. 

For smaller angles, the exact solution clearly deviates from the free solution and goes over into the Hubbard solution (green line) for $\vartheta>1.8^\circ$.  Here, the propagation efficiency is considerably reduced (cf. also Fig. \ref{fig:dynamic}). At a twist angle of 1$^\circ$, the dispersion parameter has dropped to more than one order of magnitude to $\alpha \approx 0.023~nm^2/fs$. 
 This drastic deceleration in the exciton transport velocity originates from the band flatting in the exciton dispersion (Fig.~\ref{fig:Hopping}(b)). This corresponds to trapped exciton states in real space and results in a strongly reduced group velocity $v_G=\sum_j1/\hbar(\partial E_{\mathbf{Q}}/\partial Q_j)$. Therefore, for small angles, exciton propagation is strongly slowed down \cite{brem2020tunable}. 
 \begin{figure}[t!]
    \centering
    \includegraphics[width=0.48\textwidth]{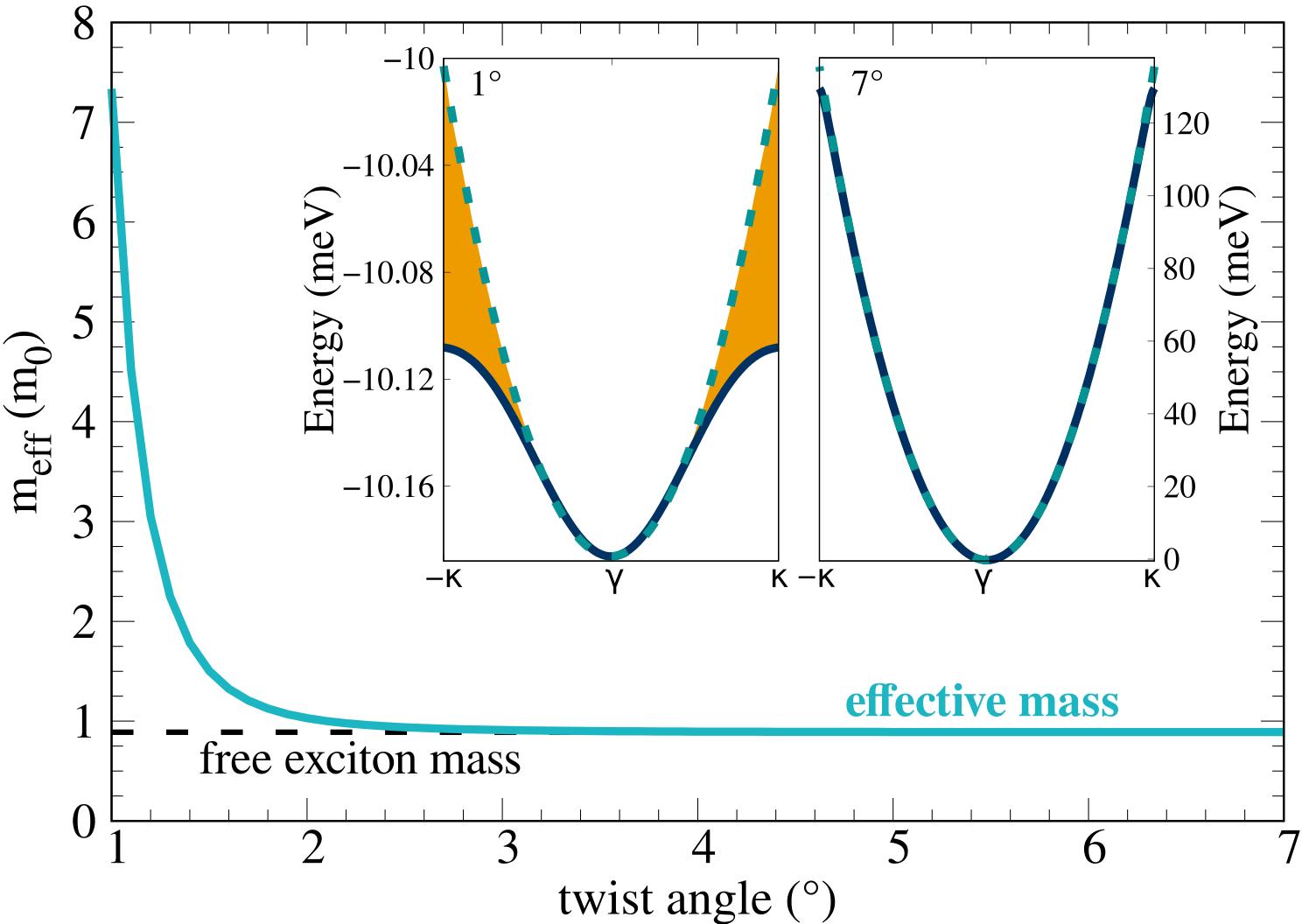}
    \caption{Effective exciton mass as a function of the twist angle $\vartheta$. For larger angles, we find that the effective mass corresponds to the free exciton mass. The insets illustrate that the parabolic approximation (dashed line) fails for small twist angles (here $1^\circ$), while it is a perfect assumption at larger angles (here $7^\circ$).
 }
    \label{fig:effective}
\end{figure}
 In this regime, we find a strongly localized exciton distribution (cf. the inset for 1$^\circ$ in Fig. \ref{fig:alpha}) and a regular continuous exciton propagation is not possible, but rather requires a hopping between different largely isolated moiré sites. Thus, we denote this as a hopping regime.  
  The Hubbard model is the most suited model to describe this hopping behavior, as one can see that the Hubbard approximation (green line) perfectly coincides with the exact solution (red line) for small twist angles $\vartheta<1.8^{\circ}$.
The dispersion parameter increases with the angle, as the overlap of exciton wavefunctions becomes larger resulting in a more efficient hopping term.   
  For larger twist angles $\vartheta>1.8^{\circ}$, we observe a clear deviation from the exact solution. The Hubbard model is described in the Wannier basis assuming that nearest neighbor hopping is the only efficient channel. This assumption is excellent for small angles with their localized states, however it fails for larger angles with strongly delocalized states. \\

\textbf{Moiré Exciton Mass.} In the presence of an external periodic potential, excitons, similar to crystal electrons, propagate as Bloch waves, which for wavepackets large compared to the cell size can be described as a change in the effective mass. These masses depend on the surrounding potential, which in the case of a monolayer is generated by the atomic lattice. In the case of heterostructures, the mass of the exciton naturally changes as a consequence of the moiré potential and therefore is expected to be very sensitive to the twist angle. Exploiting the definition of the dispersion parameter $\alpha$ in  Eq. \eqref{Eq:alpha}, we can determine an angle-dependent effective excitonic mass $m_{\text{eff}}(\vartheta)=\hbar/\alpha(\vartheta)$.  We find that for large angles $\vartheta>4^{\circ}$ the effective exciton mass converges towards the value of the the free exciton mass (with a deviation of $<1\%$ for $\vartheta>4^{\circ}$), cf. Fig. \ref{fig:effective}. However,  as $\vartheta$ decreases, we observe a drastic enhancement of the  effective mass by almost one order of magnitude at $1^\circ$. This reflects the band flattening in the excitonic band structure in presence of a moiré potential at small twist angles, cf.  Fig.~\ref{fig:Hopping}(b). In the limiting case of completely flat bands, the effective mass should approach infinity. However, for angles smaller than $1^\circ$, atomic reconstruction effects \cite{yoo2019atomic} need to be considered that are beyond the scope of this work.

\section{Summary}
We have presented a microscopic study on spatiotemporal exciton dynamics in a twisted van der Waals heterostructure exhibiting a moiré potential. We find two distinct exciton transport regimes: At large twist angles of $\vartheta>4^{\circ}$,  excitons propagate like wave packets and we denote this as dispersion regime. For smaller angles, exciton propagation is found to be strongly slowed down as the overlap of exciton wavefunctions is reduced and the exciton bands start to become flat. In the hopping regime for twist angles $\vartheta<2^{\circ}$, the Bose-Hubbard model describes accurately the propagation of excitons. Overall, we have gained new microscopic insights into the strong twist-angle dependence of exciton propagation in van der Waals heterostructures.\\

\section*{Acknowledgements}
We acknowledge funding from the Deutsche Forschungsgemeinschaft
(DFG) via SFB 1083 and the European Union’s Horizon 2020 research and innovation program under grant agreement no. 881603 (Graphene Flagship).

\bibliography{Bib}
\end{document}


\title{Exciton transport in a moiré potential: from hopping to dispersive regime\\
\Large Supplementary Information}
\author{Willy Knorr}
\email{knorrw@uni-marburg.de}
\author{Samuel Brem}
\author{Giuseppe Meneghini}
\author{Ermin Malic}
\affiliation{Department of Physics, Philipps University, 35037 Marburg, Germany}
\date{\today}
\maketitle
\linespread{1}

\subsection{Effective Single Particle Hamiltonian}
We derive an effective single-particle Hamiltonian in order to evaluate the excitonic band structure in the presence of a moiré potential. For this purpose we introduce electron-hole pair operators
$P^{\dagger}_{l\mathbf{k},l'\mathbf{k}'}=\sum_\nu X^{\nu\dagger}_{ll',\mathbf{k}-\mathbf{k}'}\psi_{ll'}^{\nu}(\alpha_{ll'}\mathbf{k'}+\beta_{ll'}\mathbf{k})$, where we expanded into excitonic eigenmodes \cite{brem2020tunable}. Here, the factors $\alpha_{ll'}/\beta_{ll'}$ include the respective electron and hole masses in the conduction/valence band. These values were taken from Ref.\cite{kormanyos2015k}. The exciton wave function fulfil the Wannier equation \cite{koch2006semiconductor}. In the low density limit and focusing on the excitonic ground state we obtain the free exciton Hamiltonian \cite{brem2020tunable}
\begin{equation}
    H_F = \sum_{ll'\textbf{Q}}\varepsilon_{\textbf{Q}}^{ll'}X^{\dagger}_{ll'\textbf{Q}}X_{ll'\textbf{Q}}.
\end{equation}
where $\varepsilon_{\textbf{Q}}^{ll'}$ denotes the free exciton dispersion. Additionally, we transform the Hamiltonian for the moiré potential \cite{brem2020tunable} reading
\begin{align}
H_M&=\sum_{ll'\textbf{Q}\textbf{q}}\text{M}_{\textbf{q}}^{ll'}X^{\dagger}_{ll'\textbf{Q}+\textbf{q}}X_{ll'\textbf{Q}}+h.c.
\end{align}
with the matrix element $\text{M}_{\textbf{q}}^{ll'} = V_l^c(\textbf{q})\mathcal{J}_{ll'}(\beta_{ll'}\textbf{q})-V_{l'}^v(\textbf{q})\mathcal{J}^*_{ll'}(\alpha_{ll'}\textbf{q})$ including the electronic moiré potential $V^{v/c}_l$ in the valence and conduction band respectively, as well as the excitonic form factor $\mathcal{J}_{ll'}(\textbf{q})=\sum_k\psi^*_{ll'}(\textbf{k})\psi_{ll'}(\textbf{k}+\textbf{q})$ \cite{brem2020tunable}. For the electronic moir\'e potentials we employ the microscopic approach developed in Ref. \cite{brem2020tunable} yielding
\begin{align}
    H_M &= \sum_{L\textbf{Q}n}\zeta_LX^{\dagger}_{L\textbf{Q}+(-1)^{l_e}\textbf{G}_n}X_{L\textbf{Q}}\quad\text{with}\\
    \zeta_L&=\begin{cases}v_{l_e}^c\mathcal{J}_{L}(\beta_L\textbf{G}_0)-v_{l_h}^v\mathcal{J}_{L}(\alpha_L\textbf{G}_0)~~\text{for}~l_e=l_h\\ v_{l_e}^c\mathcal{J}_{L}(\beta_L\textbf{G}_0)-v_{l_h}^{v*}\mathcal{J}_{L}(\alpha_L\textbf{G}_0)~~\text{for}~l_e\neq l_h\end{cases},
\end{align}
where the parameters $v_{l}^\lambda$ are extracted from first principle computations \cite{brem2020tunable}. The moiré potential creates a superlattice with a periodicity $|\mathbf{G_0}|$, which only allows the mixing between these discrete center-of-mass momenta. We utilise a zone-folding approach to take advantage of the new periodic lattice. Hence, by changing into the in the zone-folded eigenbasis $F^{\dagger}_{Ls\mathbf{Q}}=\tilde{X}_{L,\mathbf{Q}+\mathbf{G}_s}$ we obtain
 \begin{align}
     H = \sum_{Ls\textbf{Q}}\tilde{\varepsilon}_{L,\textbf{Q}+s_1\textbf{G}_1+s_2\textbf{G}_2}F^{\dagger}_{Ls\textbf{Q}}F_{Ls\textbf{Q}}+\sum_{Lss'\textbf{Q}}\tilde{\mathcal{M}}_{ss'}^{L}F^{\dagger}_{Ls\textbf{Q}}F_{Ls'\textbf{Q}}
 \end{align}
 with the modified moiré mixing matrix  
 \begin{align}
     \tilde{\text{M}}_{ss'}^{L} &= \zeta_L\Bigg(\delta(s_1,s_1'+(-1)^{l_e})\delta(s_2,s'_2)+\delta(s_1,s_1')\delta(s_2,s'_2+(-1)^{l_e})+\delta(s_1,s'_1-(-1)^{l_e})\delta(s_2,s'_2-(-1)^{l_e})\Bigg)\notag\\
     &+\zeta^*_L\bigg(\delta(s_1,s_1'-(-1)^{l_e})\delta(s_2,s'_2)+\delta(s_1,s_1')\delta(s_2,s'_2-(-1)^{l_e})+\delta(s_1,s'_1+(-1)^{l_e})\delta(s_2,s'_2+(-1)^{l_e})\bigg).
 \end{align}
We change to the eigenbasis
 \begin{align}
     Y^{\dagger}_{L\nu\textbf{Q}} = \sum_sc_{Ls}^{\nu*}(\textbf{Q})F^{\dagger}_{Ls\textbf{Q}}.
 \end{align}
 where the coefficients fulfil the eigenvalue equation
 \begin{align}
     \tilde{\varepsilon}_{L,\textbf{Q}+s_1\textbf{G}_1+s_2\textbf{G}_2}c_{Ls}^{\nu}(\textbf{Q})+\sum_{s'}\tilde{\text{M}}_{ss'}^{L}C_{Ls'}^{\nu}(\textbf{Q}) = E_{L\mu\textbf{Q}}C_{Ls}^{\nu}(\textbf{Q}).
 \end{align}
This new basis leads us to the final diagonal Hamiltonian 
 \begin{align}\label{Eq:singleHamiltonianFinal}
     H = \sum_{L\mu\textbf{Q}}E_{L\mu\textbf{Q}}Y_{L\mu\textbf{Q}}^{\dagger}Y_{L\mu\textbf{Q}}
 \end{align}
 which is further discussed in the main text.
\bibliography{Bib}